\documentclass[aps,prl,showpacs,twocolumn,superscriptaddress]{revtex4-1}
\pdfoutput=1
\usepackage{amsmath,amsthm,amssymb}
\usepackage{array}
\usepackage{graphicx}
\usepackage{fancyhdr}
\usepackage{color}
\usepackage{extarrows}
\usepackage{enumerate}
\usepackage{epstopdf}
\usepackage[colorlinks,
            linkcolor=black,
            citecolor=black,
            urlcolor=blue
            ]{hyperref}
\usepackage{natbib}
\begin{document}

\title{Exact Mapping Noisy van der Pol Type Oscillator onto Quasi-symplectic Dynamics}
\author{Ruoshi Yuan}
\affiliation{School of Biomedical Engineering}
\author{Xinan Wang}
\author{Yian Ma}
\author{Bo Yuan}
\email{Corresponding author. Email: boyuan@sjtu.edu.cn}
\affiliation{Department of Computer Science and Engineering}
\author{Ping Ao}
\email{Corresponding author. Email: aoping@sjtu.edu.cn}
\affiliation{Shanghai Center for Systems Biomedicine and Department of Physics\\Shanghai Jiao Tong University, Shanghai, 200240, China}

\date{\today}

\begin{abstract}
We find exact mappings for a class of limit cycle systems with noise onto quasi-symplectic dynamics, including a van der Pol type oscillator. A dual role potential function is obtained as a component of the quasi-symplectic dynamics. Based on a stochastic interpretation different from the traditional Ito's and Stratonovich's, we show the corresponding steady state distribution is the familiar Boltzmann-Gibbs type for arbitrary noise strength. The result provides a new angle for understanding processes without detailed balance and can be verified by experiments.
\end{abstract}
\maketitle

Noise disturbed limit cycle dynamics is now attracting considerable attention in the physics community \cite{PhysRevLett.93.204103,PhysRevLett.101.154101,*PhysRevLett.102.194102,*PhysRevE.80.036113,ge2012landscapes}. A direct reason is that ubiquitous real systems can be modeled by them, e.g., from cell cycle \cite{Tyson2003221}, circadian phenomena \cite{leloup1999limit} to chemical reaction \cite{qian2002concentration} and oscillating electrical circuit \cite{van1926lxxxviii}. The dynamics itself is a touchstone to study nonlinear dissipative process in the absence of detailed balance. Due to the difficulty arising out of nonlinearity and stochasticity, approximated methods based on phase reduction and weak noise perturbation are proposed from former studies \cite{PhysRevLett.93.204103,PhysRevLett.101.154101,*PhysRevLett.102.194102,*PhysRevE.80.036113}, but exact results are rarely seen in literature. In this Letter, we examine noised limit cycles through exactly mapping them onto quasi-symplectic \cite{milstein2003quasi} dynamics (defined in Eq.~\eqref{decomp}) valid for an arbitrary noise strength. A potential function is obtained therein as a component of the quasi-symplectic dynamics. These functions serve as both a Lyapunov function of the deterministic part dynamics \cite{lasalle1976stability} and a Hamiltonian leading to Boltzmann-Gibbs type steady state distribution of the stochastic process \cite{ao2004,*Kwon2005,*Ao2008}.

The existence of such potential function for processes without detailed balance is still suspected \cite{Strogatz2000,*carneiro2010adaptive,*eps336423,ge2012landscapes}, a specific argument is that: for a limit cycle system with non-constant velocity along the cycle, the dual role potential does not exist. We begin with such an example.
The van der Pol oscillator \cite{van1926lxxxviii} is a famous limit cycle dynamics, here we examine a stochastic version with a multiplicative noise $\zeta(\mathbf{q},t)=(\zeta_1(\mathbf{q},t),\zeta_2(\mathbf{q},t))^\tau$ (the superscript $\tau$ denotes the transpose of a matrix) and a higher order term $h(q_1)$:
\begin{figure}
  \includegraphics[width=0.45\textwidth]{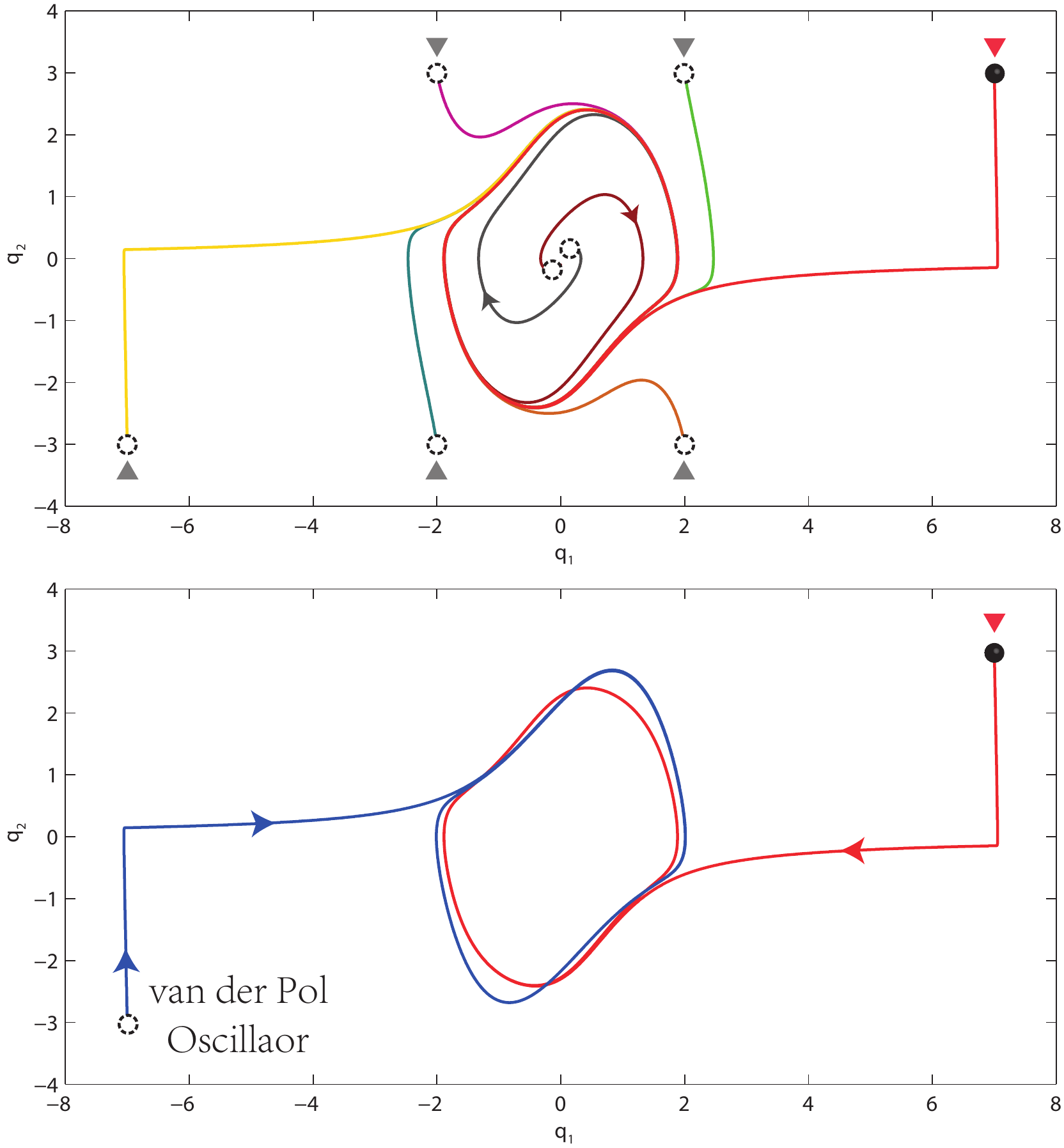}
\caption{Upper panel: Trajectories (deterministic dynamics) for the system Eq.~\eqref{eq:vanlike} with $h(q_1)=\mu^2q_1^3/4-\mu^2q_1^5/16$ ($\mu=1$). Lower panel: Comparison of two systems, the van der Pol Oscillator $h(q_1)=0$ is in blue, the system in the upper panel is in red.}
\label{fig:vanlike1}
\end{figure}
\begin{equation}
\label{eq:vanlike}
   \left\{
      \begin{array}{l}
        \dot{q}_1 = q_2 + \zeta_1(\mathbf{q},t)\\
        \dot{q}_2 = -\mu(q_1^2-1)q_2-q_1+h(q_1)+\zeta_2(\mathbf{q},t)
      \end{array}
   \right.~.
\end{equation}
When $h(q_1)=0$, the deterministic part of the dynamics reduces to the van der Pol oscillator. A specific system we would like to illustrate is $h(q_1)=\mu^2q_1^3/4-\mu^2q_1^5/16$ \cite{Abdelkader1998308}, we can observe from Fig.~\ref{fig:vanlike1}, the deterministic dynamical behavior of the system. It has a rotationally non-symmetric limit cycle and a position dependent velocity along the cycle. We propose three questions here:
\begin{enumerate}[(1).]
 \item Whether a potential function (Lyapunov function) can be explicitly constructed for the system in the upper panel of Fig.~\ref{fig:vanlike1}?
 \item Under which kind of stochastic integration can the SDEs in Eq.~\eqref{eq:vanlike} have a Boltzmann-Gibbs steady state distribution for an arbitrary noise strength?
 \item What is the significance of constructing such potential functions?
\end{enumerate}

To answer these questions, we should first briefly review a framework \cite{ao2004,*Kwon2005,*Ao2008}. Consider the Langevin equation \cite{Gardiner2004}:
\begin{align}
\label{Langevin}
\dot{\mathbf{q}}=\mathbf{f}(\mathbf{q})+N(\mathbf{q})\xi(t)~,
\end{align}
where $\mathbf{q}$, $\mathbf{f}$ are $n$-dimensional vectors and $\mathbf{f}$ is a nonlinear function of the state variable $\mathbf{q}$.
The noise term $\xi(t)$ is $k$-dimensional Gaussian white with the zero mean, $\left<\xi(t)\right>=0$, and the covariance $\left<\xi(t)\xi^\tau(t')\right>=\delta(t-t')I_k$. The notation $\delta(t-t')$ is the Dirac delta function, $\left<...\right>$ indicates the average over noise distribution, $I_k$ is the $k$-dimensional identity matrix. The element of the $n\times k$ matrix $N(\mathbf{q})$ can be a nonlinear function of $\mathbf{q}$, then the noise considered in this framework can be a general multiplicative noise. This matrix is further described by: $N(\mathbf{q})N^\tau(\mathbf{q})=2\epsilon D(\mathbf{q})$, the constant $\epsilon$ quantifying the noise strength and $D(\mathbf{q})$ is a $n\times n$ positive semi-definite diffusion matrix. Note that the noise may have less than $n$ independent components $k< n$, leading to the zero eigenvalue of $D(\mathbf{q})$.
During the study of a biological switch \cite{ao2004,*Kwon2005,*Ao2008}, a quasi-symplectic dynamics equivalent to Eq.~\eqref{Langevin} was discovered:
\begin{align}
\label{decomp}
[S(\mathbf{q})+A(\mathbf{q})]\dot{\mathbf{q}}=-\nabla\phi(\mathbf{q})+\hat{N}(\mathbf{q})\xi(t)~.
\end{align}
The term $S(\mathbf{q})$ is a positive semi-define matrix, $-S(\mathbf{q})\dot{\mathbf{q}}$ denotes a frictional force; the term $A(\mathbf{q})$ is an antisymmetric matrix representing an embedded symplectic structure, $-A(\mathbf{q})\dot{\mathbf{q}}$ is a rewritten of the Lorentz force $e \dot{\mathbf{q}}\times\mathbf{B}$ in 2 or 3 dimensional cases, and also a generalization to higher dimensions; the scalar function $\phi(\mathbf{q})$ is a potential function, e.g., the electrostatic potential, lying at the core of the discussion in this Letter. The matrix $\hat{N}(\mathbf{q})$ is constrained by the fluctuation-dissipation theorem \cite{callen1952theorem,*Kubo1966}: $\hat{N}(\mathbf{q})\hat{N}^\tau(\mathbf{q})=2\epsilon S(\mathbf{q})$.

A corresponding Fokker-Planck equation for \eqref{decomp} (therefore for Eq.~\eqref{Langevin}) can be obtained with physical significance (a zero mass limit) \cite{yuan2012beyond}:
\begin{equation}
\label{FPE}
\partial_t \rho(\mathbf{q},t)=\nabla\cdot\left[D(\mathbf{q})+Q(\mathbf{q})\right]\cdot[\nabla\phi(\mathbf{q})+\epsilon\nabla]\rho(\mathbf{q},t)~,
\end{equation}
where $\nabla$ in $\nabla\phi(\mathbf{q})$ does not operate on $\rho(\mathbf{q},t)$; $D(\mathbf{q})$ is the diffusion matrix; the matrix $Q(\mathbf{q})$ is antisymmetric and can be calculated from the relation $\left[S(\mathbf{q})+A(\mathbf{q})\right]\left[D(\mathbf{q})+Q(\mathbf{q})\right]=I_n$.
Equation \eqref{FPE} has the Boltzmann-Gibbs distribution with the potential $\phi(\mathbf{q})$ as a steady state solution:
\begin{equation}
\label{steady}
\rho(\mathbf{q},t\to\infty)=\frac{1}{Z_s}\exp\left\{\frac{\phi(\mathbf{q})}{\epsilon}\right\}
\end{equation}
where $Z_s=\int d\mathbf{q}\exp\left\{\phi(\mathbf{q})/\epsilon\right\}$ is the partition function. The 1-d case has been verified by an experiment \cite{volpe2010influence}.
The probability current density $\mathbf{j}(\mathbf{q},t)=(j_1(\mathbf{q},t),\dots,j_n(\mathbf{q},t))^\tau$ is commonly defined as:
\begin{align}
  j_i(\mathbf{q},t)=\bar f_i(\mathbf{q})\rho(\mathbf{q},t)-\partial_{j} \left[\epsilon D_{ij}(\mathbf{q})\rho(\mathbf{q},t)\right]
\end{align}
where $ \bar f_i(\mathbf{q})=f_i(\mathbf{q})+\epsilon\partial_j[D_{ij}(\mathbf{q})+Q_{ij}(\mathbf{q})]$, $f_i(\mathbf{q})$ is the $i$th component of the vector valued function $\mathbf{f}(\mathbf{q})$ in Eq.~\eqref{Langevin},
 $D_{ij}(\mathbf{q})$ and $Q_{ij}(\mathbf{q})$ are the elements of the matrices
 $D(\mathbf{q})$ and $Q(\mathbf{q})$ in Eq.~\eqref{FPE}.
 In steady state, the probability distribution is given by Eq.~\eqref{steady}. We have $\nabla\cdot\mathbf{j}_s(\mathbf{q})=0$, but $\mathbf{j}_s(\mathbf{q})$ is usually not zero.
One can check that $Q=0$ is a sufficient condition for $\mathbf{j}_s=0$;
but when $Q(\mathbf{q})\neq 0$ then generally $\mathbf{j}_s(\mathbf{q})\neq0$,
since $\partial_j\left[Q_{ij}(\mathbf{q})\rho_s(\mathbf{q})\right]\neq0$. Therefore the framework encompasses the cases without detailed balance.
The term ``detailed balance" means the net current between any two states in the phase space is zero \cite{Kubo1995},
identical to that for Markov process in mathematics. The dynamics studied in this Letter corresponds to the non-detailed balance cases discussed in \cite{san1980limit} as well.

Equation \eqref{FPE} then defines a stochastic interpretation for the Langevin equation Eq.~\eqref{Langevin}, that is the A-type integration.
The relation between friction $S(\mathbf{q})$ and diffusion $D(\mathbf{q})$ in the absence of detailed balance condition is named as the generalized Einstein relation in \cite{ao2004,*Kwon2005,*Ao2008}, in 1-d case, it reduces to the well-known Einstein relation.
A detailed derivation of Eq.~\eqref{FPE} has been provided in \cite{yuan2012beyond}.

The Langevin equation~\eqref{Langevin} can also be considered as a composition of a \textit{deterministic dynamics} $\dot{\mathbf{q}}=\mathbf{f}(\mathbf{q})$ and a fluctuation $N(\mathbf{q})\xi(t)$.
For the deterministic dynamics, the time derivative of the potential function $\phi(\mathbf{q})$ along a trajectory is:
\begin{align*}
\frac{d\phi(\mathbf{q})}{dt}&=\nabla \phi \cdot\mathbf{f}(\mathbf{q})=-\nabla\phi(\mathbf{q})\cdot\left[D(\mathbf{q})+Q(\mathbf{q})\right]\cdot\nabla\phi(\mathbf{q})\nonumber\\
&=-\nabla\phi(\mathbf{q})\cdot D(\mathbf{q})\cdot\nabla\phi(\mathbf{q})\leq 0~,
\end{align*}
since $D(\mathbf{q})$ is nonnegative and symmetric. It means that the potential along the trajectory is non-increasing and has the local extreme values at fixed points, limit cycles or more complex attractors. Hence, potential function $\phi(\mathbf{q})$ servers as a Lyapunov function \cite{lasalle1976stability} for the deterministic dynamics $\dot{\mathbf{q}}=\mathbf{f}(\mathbf{q})$. When noise exists, the system can deviate from the attractors, transferring from one stable state to another, once the system enters a new basin of attraction, the deterministic force (dissipation) will drive the system to the corresponding attractor. It shows that noise is sometimes essential in mathematical modeling of reality and plays a driving role in phase transitions.

For all planar rotationally symmetric limit cycle dynamics (represented below in polar coordinate $(q=\sqrt{q_1^2+q_2^2},\theta)$ for simplicity) with the diffusion matrix $D=D_0I_2$ ($D_0$ is the constant diffusion coefficient):
\begin{align}
\label{polar_sys}
 \left\{\begin{array}{c}
   \dot{q}=R(q)+\zeta_q(q,\theta,t)\\
   \dot{\theta}=\psi(q)+\zeta_\theta(q,\theta,t)
 \end{array}\right.~,
\end{align}
we can provide the exact construction for the potential function (note that the following result should be transformed back to Cartesian coordinate):
\begin{align}
 \phi(\mathbf{q})=-\frac{1}{D_0}\int R(q) dq~,
\end{align}
and corresponding dynamical components:
\begin{align}
    S(\mathbf{q})&=\frac{R(q)^2}{D_0\left[R(q)^2+q^2\psi(q)^2\right]}\left(
      \begin{array}{cc}
        1 & 0\\
        0 & 1
      \end{array}
    \right)~, \nonumber\\
    A(\mathbf{q})&=\frac{q\psi(q)R(q)}{D_0\left[R(q)^2+q^2\psi(q)^2\right]}\left(
      \begin{array}{cc}
        0 & 1\\
        -1 & 0
      \end{array}
      \right)~,\nonumber\\
    Q(\mathbf{q})&=-\frac{q\psi(q)D_0}{R(q)}\left(
      \begin{array}{cc}
        0 & 1\\
        -1 & 0
      \end{array}
      \right)~.
\end{align}
Note that for the weak noise limit, when $D_0\to0$, this construction is still valid for the deterministic dynamics by an arbitrarily setting $D_0$ (see also \cite{Yuan2010}).

More generally, we can extend our method to rotationally non-symmetric systems with multiplicative noise through coordinate transformations (reversible continuous mapping, can be nonlinear) $\sigma$ of the deterministic dynamics:
\begin{align*}
\left\{\begin{array}{l}
    \dot{q}_1=f_1(\mathbf{q}) \\
    \dot{q}_2=f_2(\mathbf{q})
    \end{array}
    \right. \xlongleftrightarrow[\sigma^{-1}]{\sigma}
    \left\{\begin{array}{l}
    \dot{u}=\bar f_1(u,v) \\
    \dot{v}=\bar f_2(u,v)
    \end{array}
    \right.
    \leftrightarrow
\left\{
\begin{array}{l}
\dot{q}=\rho(q)P(q,\theta)\\
\dot{\theta}=\varphi(q,\theta)
\end{array}
\right.
\end{align*}
where $P(q,\theta)\geqslant0$. The potential function is given by
\begin{equation*}
\phi(q_1,q_2)\xlongleftarrow{\sigma^{-1}(u,v)}\bar\phi(u,v)\xlongleftarrow{polar^{-1}(q)}\bar\phi_p(q)=-\int{\rho(q)dq}~,
\end{equation*}
since $d\phi(q_1,q_2)/dt=\left[d\bar\phi_p(q)/dq\right]\dot{q}=-\rho^2(q)P(q,\theta)\leq 0$.
The protocol of the construction: First, for the deterministic dynamics $(q_1,q_2)$, find out the transformation $\sigma: (q_1,q_2)\to(u,v)$, calculate the dynamics under $(u,v)$, that is the $\dot{u}=\bar f_1(u,v)$ and $\dot{v}=\bar f_2(u,v)$; Second, rewrite the obtained dynamics in polar coordinates $(u,v)\to(q,\theta)$, if the dynamics can be expressed as the requested form above, the potential function can be constructed as $\bar\phi_p(q)$; Third, transform $\bar\phi_p(q)$ back to $\bar\phi(u,v)$, and finally to $\phi(q_1,q_2)$.

Once potential function (Lyapunov function) $\phi(\mathbf{q})$ for the deterministic dynamics $\dot{\mathbf{q}}=\mathbf{f}(\mathbf{q})$ is obtained, there are different ways to construct the dynamical components, one particular setting is provided in \cite{Yuan2010} (the binary operator of two $n$ dimensional vectors is defined as $\mathbf{x}\times\mathbf{y}=(x_iy_j-x_jy_i)_{n\times n}$, the result is an $n\times n$ matrix):
\begin{align}
\label{construction}
S(\mathbf{q})&=-\frac{\nabla \phi\cdot\mathbf{f}}{\mathbf{f}\cdot\mathbf{f}}I,\quad A(\mathbf{q})=-\frac{\nabla \phi\times \mathbf{f}}{\mathbf{f}\cdot\mathbf{f}}~,\nonumber\\
D(\mathbf{q})&=-\left[\frac{\mathbf{f}\cdot\mathbf{f}}{\nabla \phi \cdot \mathbf{f}}
I+\frac{\left(\nabla \phi\times \mathbf{f}\right)^2}{\left(\nabla \phi \cdot \mathbf{f}\right)\left(\nabla \phi\cdot\nabla \phi\right)}\right]~,\nonumber\\
Q(\mathbf{q})&=\frac{\nabla \phi\times \mathbf{f}}{\nabla \phi\cdot\nabla \phi}~.
\end{align}
We should note that although the potential function $\phi$ is invariant during coordinate transformation, the dynamical components, the matrices $S$, $A$, $D$, $Q$ vary in different coordinates, but a straightforward formulation can be achieved by multiplying Jacobian matrix of the transformation.

\begin{figure}[!]
\includegraphics[width=0.45\textwidth]{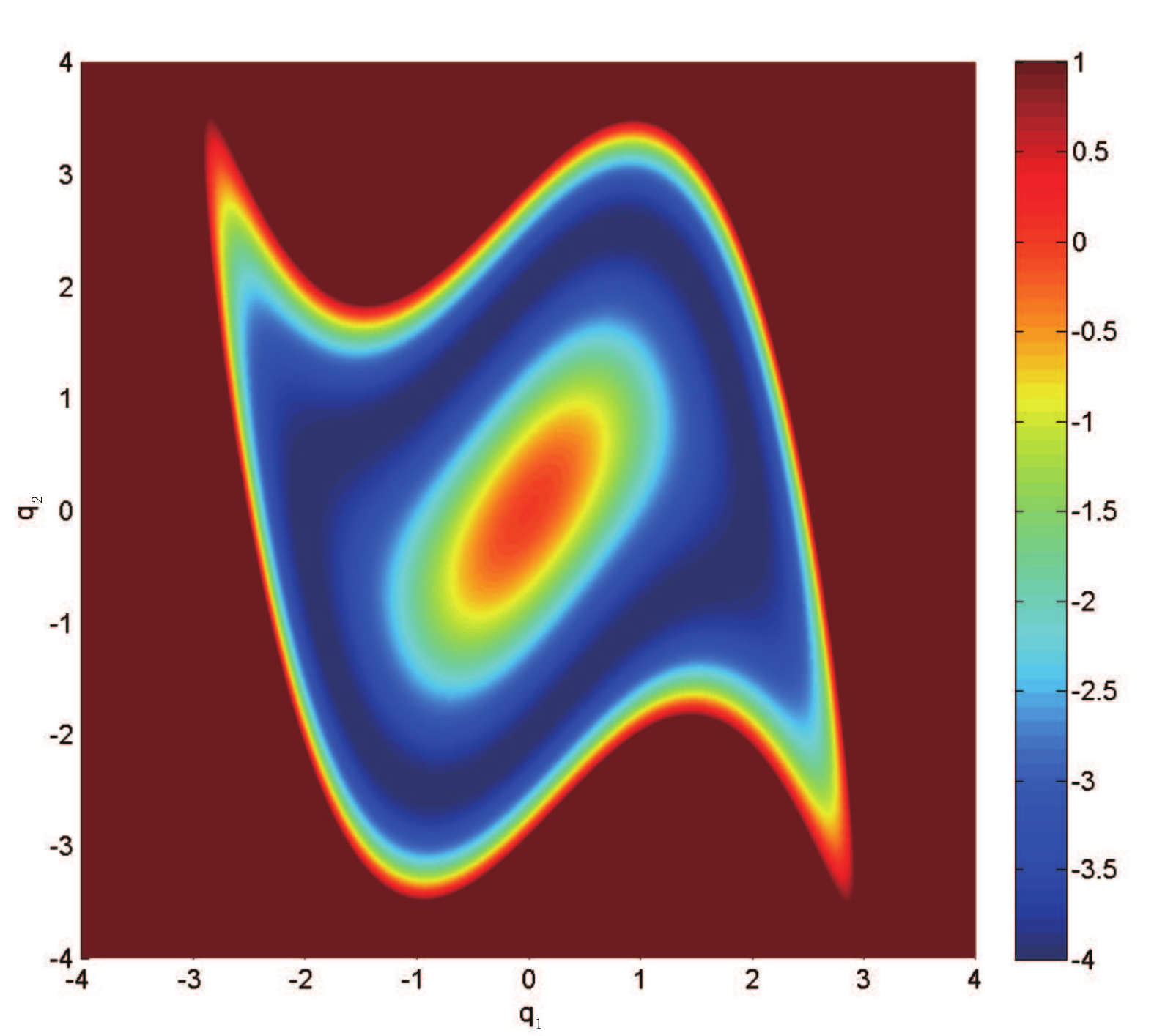}
\caption{Potential function Eq.~\eqref{vanlike} with $\mu=1$: The color blue denotes a lower potential value, and the red means higher value. The graph is drawn below a preset upper bound value $1$, the phase variables are $q_1$ and $q_2$.}
\label{fig:vanlike}
\end{figure}

Back to the example in Eq.~\eqref{eq:vanlike} with $(\zeta_1(\mathbf{q},t),\zeta_2(\mathbf{q},t))^\tau$ $=N(\mathbf{q})\cdot(\xi_1(t),\xi_2(t))^\tau$ and $h(q_1)=\mu^2q_1^3/4-\mu^2q_1^5/16$. The deterministic part is a Li\'enard equation similar to the famous van der Pol Oscillator ($0<\mu<2$, see Fig.~\ref{fig:vanlike}) \cite{Abdelkader1998308}.
Through a nonlinear coordinate transformation $\sigma^{-1}$: $u=q_1$ and $v=q_2-\mu q_1+\mu q_1^3/4$, we obtain the dynamical system:
\begin{equation*}
\left\{
\begin{array}{cl}
\dot{u}&=\mu u-\frac{\mu}{4}u^3+v\\
\dot{v}&=-u-\frac{\mu}{4} u^2v
\end{array}
\right.\leftrightarrow\left\{
\begin{array}{cl}
 \dot{q}&=\frac{\mu}{4}\left(4-q^2\right)q\cos^2\theta\\
 \dot{\theta}&=-1-\mu\cos\theta\sin\theta
 \end{array}
\right.,
\end{equation*}
with $\rho(q)=(4-q^2)q$ and $P(q,\theta)=\mu \cos^2\theta/4\geq0$ ($\mu>0$). Therefore, we can provide an exact construction of potential function for Eq.~\eqref{eq:vanlike} (see Fig.~\ref{fig:vanlike}):
\begin{align}
\label{vanlike}
\phi(\mathbf{q})=\frac{1}{4}\left[q_1^2+\left(q_2-\mu q_1+\frac{\mu}{4}q_1^3\right)^2\right]\times\nonumber\\
\left[q_1^2+\left(q_2-\mu q_1+\frac{\mu}{4}q_1^3\right)^2-8\right]~.
\end{align}
We note that the potential function Eq.~\eqref{vanlike} has the minimal value at the stable limit cycle
$\dot{q_1}=\mu q_1-\frac{\mu}{4}q_1^3 \pm \sqrt{4-q_1^2}$
 and a local maximum value at the unstable fixed point $(0,0)$.
Expressions for other dynamical components can be provided through Eq.~\eqref{construction}. We use the representation with $(u=q_1,v=\dot{q_2-\mu q_1+\mu q_1^3/4})$ and $J(\mathbf{q})$ the Jacobian matrix $\partial(u,v)/\partial(q_1,q_2)$:
\begin{widetext}
\begin{align}
  \label{diffusion}
  S(\mathbf{q})&=\frac{\mu(4-u^2-v^2)^2 u^2}{4\left(\dot{u}^2+\dot{v}^2\right)}J(\mathbf{q})^{\tau}J(\mathbf{q})~,~
  A(\mathbf{q})=-\frac{(4-u^2-v^2)(u^2+v^2+\mu uv)}{\dot{u}^2+\dot{v}^2} J(\mathbf{q})^{\tau}\left(
      \begin{array}{cc}
        0 & 1\\
        -1 & 0
      \end{array}
      \right)J(\mathbf{q})~,\nonumber\\
  D(\mathbf{q})&=\frac{\mu u^2}{4\left(u^2+v^2\right)}J(\mathbf{q})^{-1}J(\mathbf{q})^{-\tau}~,~
  Q(\mathbf{q})=\frac{u^2+v^2+\mu uv}{(u^2+v^2)(4-u^2-v^2)}J(\mathbf{q})^{-1}\left(
      \begin{array}{cc}
        0 & 1\\
        -1 & 0
      \end{array}
      \right)J(\mathbf{q})^{-\tau}~.
\end{align}
\end{widetext}
The result obtained can be understand as: The stochastic dynamics Eq.~\eqref{eq:vanlike} with the position dependent diffusion matrix $D(\mathbf{q})$ given in Eq.~\eqref{diffusion} has the explicitly constructed potential function $\phi(\mathbf{q})$ (Eq.~\eqref{vanlike}) and a corresponding Boltzmann-Gibbs steady state distribution (Eq.~\eqref{steady}). For the matrix $Q(\mathbf{q})$, one can check $\partial_j\left[Q_{ij}(\mathbf{q})\rho_s(\mathbf{q})\right]\neq0$, leading to the absence of detailed balance.
The stochastic integration used is the A-type (see Eq.~\eqref{FPE}) different from traditional Ito's or Stratonovich's \cite{yuan2012beyond}.
Note that when approaching to the limit cycle $(4-u^2-v^2)\to 0$, the force induced by the potential gradient goes to zero; the Lorentz force matrix $A(\mathbf{q})$ goes zero in the same order and changes its sign at the limit cycle (since $0<\mu<2$); the friction matrix $S(\mathbf{q})$ goes to zero in a higher order.
The dynamics at the limit cycle is no longer dissipative but conserved in this limit, reaching a stable cycle. Thus the potential value should be equal on limit cycles where the system is conserved. We note that this is consistent with the definition of a Lyapunov function \cite{lasalle1976stability}, since the particle is moving repeatedly along the cycle, the same as the conserved system moving along the Hamiltonian. The singularity problem for this construction has been discussed in \cite{Yuan2010}. Previous works focus more on the diffusion matrix, ignoring the important role played by the friction matrix $S(\mathbf{q})$ and the Lorentz force matrix $A(\mathbf{q})$.

The significance of constructing potential functions defined in this framework can be viewed from two aspects: First, there is a simple and direct correspondence between stochastic and determinist dynamics for arbitrary noise strength (not only under weak noise limit). For example, attractors like fixed points or limit cycles do not change their position in the phase space. This enables a straightforward use of the dynamical analysis for the deterministic dynamics in the presence of noise. Therefore, the calculation of the transition probability from one stable fixed point $\mathbf{q}_1$ to another one through a saddle point $\mathbf{q}_2$ is generally formulated as $\propto\exp\left[-|\phi(\mathbf{q}_1)-\phi(\mathbf{q}_2)|/\epsilon\right]$; Second, potential function obtained here serves also a Lyapunov function for the deterministic dynamics. The framework then contributes possible new approaches for the largely unsolved problem in engineering: constructing Lyapunov function for general nonlinear dynamics. In addition, the A-type stochastic integration can be applied directly in the study of phase reduction. For conventional phase reduction method \cite{PhysRevLett.93.204103}, A-type integration does not lead to the noise-induced frequency shift (NIFS).

In conclusion, we have exactly mapped a class of limit cycle systems with noise onto quasi-symplectic dynamics, from rotationally symmetric to non-symmetric systems, a specific example is a van der Pol type oscillator. These systems have been analyzed through the dynamical components $S(\mathbf{q})$, $A(\mathbf{q})$ and $\phi(\mathbf{q})$ obtained by the mapping. Using A-type integration, the steady state distribution of these systems is Boltzmann-Gibbs type, consistent with the knowledge from statistical physics. Since new measuring techniques for Brownian motion is available (e.g. \cite{volpe2010influence}), the theoretical results here may be verified by experiments directly in the near future.

\begin{acknowledgments}
This work was supported in part by the National 973 Project No.~2010CB529200; and by the Natural Science Foundation of China No.~NFSC61073087 and No.~NFSC91029738.
\end{acknowledgments}
\bibliography{Yrs}

\begin{thebibliography}{26}%
\makeatletter
\providecommand \@ifxundefined [1]{%
 \@ifx{#1\undefined}
}%
\providecommand \@ifnum [1]{%
 \ifnum #1\expandafter \@firstoftwo
 \else \expandafter \@secondoftwo
 \fi
}%
\providecommand \@ifx [1]{%
 \ifx #1\expandafter \@firstoftwo
 \else \expandafter \@secondoftwo
 \fi
}%
\providecommand \natexlab [1]{#1}%
\providecommand \enquote  [1]{``#1''}%
\providecommand \bibnamefont  [1]{#1}%
\providecommand \bibfnamefont [1]{#1}%
\providecommand \citenamefont [1]{#1}%
\providecommand \href@noop [0]{\@secondoftwo}%
\providecommand \href [0]{\begingroup \@sanitize@url \@href}%
\providecommand \@href[1]{\@@startlink{#1}\@@href}%
\providecommand \@@href[1]{\endgroup#1\@@endlink}%
\providecommand \@sanitize@url [0]{\catcode `\\12\catcode `\$12\catcode
  `\&12\catcode `\#12\catcode `\^12\catcode `\_12\catcode `\%12\relax}%
\providecommand \@@startlink[1]{}%
\providecommand \@@endlink[0]{}%
\providecommand \url  [0]{\begingroup\@sanitize@url \@url }%
\providecommand \@url [1]{\endgroup\@href {#1}{\urlprefix }}%
\providecommand \urlprefix  [0]{URL }%
\providecommand \Eprint [0]{\href }%
\providecommand \doibase [0]{http://dx.doi.org/}%
\providecommand \selectlanguage [0]{\@gobble}%
\providecommand \bibinfo  [0]{\@secondoftwo}%
\providecommand \bibfield  [0]{\@secondoftwo}%
\providecommand \translation [1]{[#1]}%
\providecommand \BibitemOpen [0]{}%
\providecommand \bibitemStop [0]{}%
\providecommand \bibitemNoStop [0]{.\EOS\space}%
\providecommand \EOS [0]{\spacefactor3000\relax}%
\providecommand \BibitemShut  [1]{\csname bibitem#1\endcsname}%
\let\auto@bib@innerbib\@empty
\bibitem [{\citenamefont {Teramae}\ and\ \citenamefont
  {Tanaka}(2004)}]{PhysRevLett.93.204103}%
  \BibitemOpen
  \bibfield  {author} {\bibinfo {author} {\bibfnamefont {J.-n.}\ \bibnamefont
  {Teramae}}\ and\ \bibinfo {author} {\bibfnamefont {D.}~\bibnamefont
  {Tanaka}},\ }\href {\doibase 10.1103/PhysRevLett.93.204103} {\bibfield
  {journal} {\bibinfo  {journal} {Phys. Rev. Lett.}\ }\textbf {\bibinfo
  {volume} {93}},\ \bibinfo {pages} {204103} (\bibinfo {year}
  {2004})}\BibitemShut {NoStop}%
\bibitem [{\citenamefont {Yoshimura}\ and\ \citenamefont
  {Arai}(2008)}]{PhysRevLett.101.154101}%
  \BibitemOpen
  \bibfield  {author} {\bibinfo {author} {\bibfnamefont {K.}~\bibnamefont
  {Yoshimura}}\ and\ \bibinfo {author} {\bibfnamefont {K.}~\bibnamefont
  {Arai}},\ }\href {\doibase 10.1103/PhysRevLett.101.154101} {\bibfield
  {journal} {\bibinfo  {journal} {Phys. Rev. Lett.}\ }\textbf {\bibinfo
  {volume} {101}},\ \bibinfo {pages} {154101} (\bibinfo {year}
  {2008})}\BibitemShut {NoStop}%
\bibitem [{\citenamefont {Teramae}\ \emph {et~al.}(2009)\citenamefont {Teramae}
  \emph {et~al.}}]{PhysRevLett.102.194102}%
  \BibitemOpen
  \bibfield  {author} {\bibinfo {author} {\bibfnamefont {J.-n.}\ \bibnamefont
  {Teramae}} \emph {et~al.},\ }\href {\doibase 10.1103/PhysRevLett.102.194102}
  {\bibfield  {journal} {\bibinfo  {journal} {Phys. Rev. Lett.}\ }\textbf
  {\bibinfo {volume} {102}},\ \bibinfo {pages} {194102} (\bibinfo {year}
  {2009})}\BibitemShut {NoStop}%
\bibitem [{\citenamefont {Gal\'an}(2009)}]{PhysRevE.80.036113}%
  \BibitemOpen
  \bibfield  {author} {\bibinfo {author} {\bibfnamefont {R.~F.}\ \bibnamefont
  {Gal\'an}},\ }\href {\doibase 10.1103/PhysRevE.80.036113} {\bibfield
  {journal} {\bibinfo  {journal} {Phys. Rev. E}\ }\textbf {\bibinfo {volume}
  {80}},\ \bibinfo {pages} {036113} (\bibinfo {year} {2009})}\BibitemShut
  {NoStop}%
\bibitem [{\citenamefont {Ge}\ and\ \citenamefont
  {Qian}(2012)}]{ge2012landscapes}%
  \BibitemOpen
  \bibfield  {author} {\bibinfo {author} {\bibfnamefont {H.}~\bibnamefont
  {Ge}}\ and\ \bibinfo {author} {\bibfnamefont {H.}~\bibnamefont {Qian}},\
  }\href {\doibase 10.1063/1.4729137} {\bibfield  {journal} {\bibinfo
  {journal} {Chaos}\ }\textbf {\bibinfo {volume} {22}},\ \bibinfo {pages}
  {023140} (\bibinfo {year} {2012})}\BibitemShut {NoStop}%
\bibitem [{\citenamefont {Tyson}\ \emph {et~al.}(2003)\citenamefont {Tyson}
  \emph {et~al.}}]{Tyson2003221}%
  \BibitemOpen
  \bibfield  {author} {\bibinfo {author} {\bibfnamefont {J.~J.}\ \bibnamefont
  {Tyson}} \emph {et~al.},\ }\href {\doibase 10.1016/S0955-0674(03)00017-6}
  {\bibfield  {journal} {\bibinfo  {journal} {Curr. Opin. Cell Biol.}\ }\textbf
  {\bibinfo {volume} {15}},\ \bibinfo {pages} {221} (\bibinfo {year}
  {2003})}\BibitemShut {NoStop}%
\bibitem [{\citenamefont {Leloup}\ \emph {et~al.}(1999)\citenamefont {Leloup}
  \emph {et~al.}}]{leloup1999limit}%
  \BibitemOpen
  \bibfield  {author} {\bibinfo {author} {\bibfnamefont {J.~C.}\ \bibnamefont
  {Leloup}} \emph {et~al.},\ }\href {\doibase 10.1177/074873099129000948}
  {\bibfield  {journal} {\bibinfo  {journal} {J. Biol. Rhythms}\ }\textbf
  {\bibinfo {volume} {14}},\ \bibinfo {pages} {433} (\bibinfo {year}
  {1999})}\BibitemShut {NoStop}%
\bibitem [{\citenamefont {Qian}\ \emph {et~al.}(2002)\citenamefont {Qian} \emph
  {et~al.}}]{qian2002concentration}%
  \BibitemOpen
  \bibfield  {author} {\bibinfo {author} {\bibfnamefont {H.}~\bibnamefont
  {Qian}} \emph {et~al.},\ }\href {\doibase 10.1073/pnas.152007599} {\bibfield
  {journal} {\bibinfo  {journal} {Proc. Natl. Acad. Sci. USA}\ }\textbf
  {\bibinfo {volume} {99}},\ \bibinfo {pages} {10376} (\bibinfo {year}
  {2002})}\BibitemShut {NoStop}%
\bibitem [{\citenamefont {van~der Pol}(1926)}]{van1926lxxxviii}%
  \BibitemOpen
  \bibfield  {author} {\bibinfo {author} {\bibfnamefont {B.}~\bibnamefont
  {van~der Pol}},\ }\href {\doibase 10.1080/14786442608564127} {\bibfield
  {journal} {\bibinfo  {journal} {Phil. Mag.}\ }\textbf {\bibinfo {volume}
  {2}},\ \bibinfo {pages} {978} (\bibinfo {year} {1926})}\BibitemShut {NoStop}%
\bibitem [{\citenamefont {Milstein}\ and\ \citenamefont
  {Tretyakov}(2003)}]{milstein2003quasi}%
  \BibitemOpen
  \bibfield  {author} {\bibinfo {author} {\bibfnamefont {G.~N.}\ \bibnamefont
  {Milstein}}\ and\ \bibinfo {author} {\bibfnamefont {M.~V.}\ \bibnamefont
  {Tretyakov}},\ }\href {\doibase 10.1093/imanum/23.4.593} {\bibfield
  {journal} {\bibinfo  {journal} {IMA J. Numer. Anal.}\ }\textbf {\bibinfo
  {volume} {23}},\ \bibinfo {pages} {593} (\bibinfo {year} {2003})}\BibitemShut
  {NoStop}%
\bibitem [{\citenamefont {{La Salle}}(1976)}]{lasalle1976stability}%
  \BibitemOpen
  \bibfield  {author} {\bibinfo {author} {\bibfnamefont {J.~P.}\ \bibnamefont
  {{La Salle}}},\ }\href@noop {} {\emph {\bibinfo {title} {{The stability of
  dynamical systems}}}},\ \bibinfo {number} {25}\ (\bibinfo  {publisher}
  {SIAM},\ \bibinfo {address} {Philadelphia},\ \bibinfo {year}
  {1976})\BibitemShut {NoStop}%
\bibitem [{\citenamefont {Ao}(2004)}]{ao2004}%
  \BibitemOpen
  \bibfield  {author} {\bibinfo {author} {\bibfnamefont {P.}~\bibnamefont
  {Ao}},\ }\href {\doibase 10.1088/0305-4470/37/3/L01} {\bibfield  {journal}
  {\bibinfo  {journal} {J. Phys. A: Math. Gen.}\ }\textbf {\bibinfo {volume}
  {37}},\ \bibinfo {pages} {L25} (\bibinfo {year} {2004})}\BibitemShut
  {NoStop}%
\bibitem [{\citenamefont {Kwon}\ \emph {et~al.}(2005)\citenamefont {Kwon} \emph
  {et~al.}}]{Kwon2005}%
  \BibitemOpen
  \bibfield  {author} {\bibinfo {author} {\bibfnamefont {C.}~\bibnamefont
  {Kwon}} \emph {et~al.},\ }\href {\doibase 10.1073/pnas.0506347102} {\bibfield
   {journal} {\bibinfo  {journal} {Proc. Natl. Acad. Sci. USA}\ }\textbf
  {\bibinfo {volume} {102}},\ \bibinfo {pages} {13029} (\bibinfo {year}
  {2005})}\BibitemShut {NoStop}%
\bibitem [{\citenamefont {Ao}(2008)}]{Ao2008}%
  \BibitemOpen
  \bibfield  {author} {\bibinfo {author} {\bibfnamefont {P.}~\bibnamefont
  {Ao}},\ }\href {\doibase 10.1088/0253-6102/49/5/01} {\bibfield  {journal}
  {\bibinfo  {journal} {Commun. Theor. Phys.}\ }\textbf {\bibinfo {volume}
  {49}},\ \bibinfo {pages} {1073} (\bibinfo {year} {2008})}\BibitemShut
  {NoStop}%
\bibitem [{\citenamefont {Strogatz}(1994)}]{Strogatz2000}%
  \BibitemOpen
  \bibfield  {author} {\bibinfo {author} {\bibfnamefont {S.~H.}\ \bibnamefont
  {Strogatz}},\ }\href@noop {} {\emph {\bibinfo {title} {Nonlinear dynamics and
  chaos: With applications to physics, biology, chemistry, and engineering}}}\
  (\bibinfo  {publisher} {Perseus Books},\ \bibinfo {address} {Reading},\
  \bibinfo {year} {1994})\ p.\ \bibinfo {pages} {201}\BibitemShut {NoStop}%
\bibitem [{\citenamefont {Carneiro}\ and\ \citenamefont
  {Hartl}(2010)}]{carneiro2010adaptive}%
  \BibitemOpen
  \bibfield  {author} {\bibinfo {author} {\bibfnamefont {M.}~\bibnamefont
  {Carneiro}}\ and\ \bibinfo {author} {\bibfnamefont {D.~L.}\ \bibnamefont
  {Hartl}},\ }\href {\doibase 10.1073/pnas.0906192106} {\bibfield  {journal}
  {\bibinfo  {journal} {Proc. Natl. Acad. Sci. USA}\ }\textbf {\bibinfo
  {volume} {107}},\ \bibinfo {pages} {1747} (\bibinfo {year}
  {2010})}\BibitemShut {NoStop}%
\bibitem [{\citenamefont {Weinreich}\ \emph {et~al.}(2012)\citenamefont
  {Weinreich} \emph {et~al.}}]{eps336423}%
  \BibitemOpen
  \bibfield  {author} {\bibinfo {author} {\bibfnamefont {D.~M.}\ \bibnamefont
  {Weinreich}} \emph {et~al.},\ }\href {http://eprints.soton.ac.uk/336423/}
  {\bibfield  {journal} {\bibinfo  {journal} {J. Stat. Mech.}\ }\textbf
  {\bibinfo {volume} {2012}},\ \bibinfo {pages} {523} (\bibinfo {year}
  {2012})}\BibitemShut {NoStop}%
\bibitem [{\citenamefont {Abdelkader}(1998)}]{Abdelkader1998308}%
  \BibitemOpen
  \bibfield  {author} {\bibinfo {author} {\bibfnamefont {M.~A.}\ \bibnamefont
  {Abdelkader}},\ }\href {\doibase DOI: 10.1006/jmaa.1997.5746} {\bibfield
  {journal} {\bibinfo  {journal} {J. Math. Anal. Appl.}\ }\textbf {\bibinfo
  {volume} {218}},\ \bibinfo {pages} {308} (\bibinfo {year}
  {1998})}\BibitemShut {NoStop}%
\bibitem [{\citenamefont {Gardiner}(2004)}]{Gardiner2004}%
  \BibitemOpen
  \bibfield  {author} {\bibinfo {author} {\bibfnamefont {C.~W.}\ \bibnamefont
  {Gardiner}},\ }\href@noop {} {\emph {\bibinfo {title} {Handbook of stochastic
  methods: For physics, chemistry and the natural sciences}}},\ \bibinfo
  {edition} {3rd}\ ed.\ (\bibinfo  {publisher} {Springer-Verlag},\ \bibinfo
  {address} {Berlin},\ \bibinfo {year} {2004})\BibitemShut {NoStop}%
\bibitem [{\citenamefont {Callen}\ and\ \citenamefont
  {Greene}(1952)}]{callen1952theorem}%
  \BibitemOpen
  \bibfield  {author} {\bibinfo {author} {\bibfnamefont {H.~B.}\ \bibnamefont
  {Callen}}\ and\ \bibinfo {author} {\bibfnamefont {R.~F.}\ \bibnamefont
  {Greene}},\ }\href {\doibase 10.1103/PhysRev.86.702} {\bibfield  {journal}
  {\bibinfo  {journal} {Phys. Rev.}\ }\textbf {\bibinfo {volume} {86}},\
  \bibinfo {pages} {702} (\bibinfo {year} {1952})}\BibitemShut {NoStop}%
\bibitem [{\citenamefont {Kubo}(1966)}]{Kubo1966}%
  \BibitemOpen
  \bibfield  {author} {\bibinfo {author} {\bibfnamefont {R.}~\bibnamefont
  {Kubo}},\ }\href {\doibase doi:10.1088/0034-4885/29/1/306} {\bibfield
  {journal} {\bibinfo  {journal} {Rep. Prog. Phys.}\ }\textbf {\bibinfo
  {volume} {29}},\ \bibinfo {pages} {255} (\bibinfo {year} {1966})}\BibitemShut
  {NoStop}%
\bibitem [{\citenamefont {Yuan}\ and\ \citenamefont
  {Ao}(2012)}]{yuan2012beyond}%
  \BibitemOpen
  \bibfield  {author} {\bibinfo {author} {\bibfnamefont {R.-S.}\ \bibnamefont
  {Yuan}}\ and\ \bibinfo {author} {\bibfnamefont {P.}~\bibnamefont {Ao}},\
  }\href {\doibase doi:10.1088/1742-5468/2012/07/P07010} {\bibfield  {journal}
  {\bibinfo  {journal} {J. Stat. Mech.}\ }\textbf {\bibinfo {volume} {2012}},\
  \bibinfo {pages} {P07010} (\bibinfo {year} {2012})}\BibitemShut {NoStop}%
\bibitem [{\citenamefont {Volpe}\ \emph {et~al.}(2010)\citenamefont {Volpe}
  \emph {et~al.}}]{volpe2010influence}%
  \BibitemOpen
  \bibfield  {author} {\bibinfo {author} {\bibfnamefont {G.}~\bibnamefont
  {Volpe}} \emph {et~al.},\ }\href {\doibase 10.1103/PhysRevLett.104.170602}
  {\bibfield  {journal} {\bibinfo  {journal} {Phys. Rev. Lett.}\ }\textbf
  {\bibinfo {volume} {104}},\ \bibinfo {pages} {170602} (\bibinfo {year}
  {2010})}\BibitemShut {NoStop}%
\bibitem [{\citenamefont {Kubo}\ \emph {et~al.}(1995)\citenamefont {Kubo} \emph
  {et~al.}}]{Kubo1995}%
  \BibitemOpen
  \bibfield  {author} {\bibinfo {author} {\bibfnamefont {R.}~\bibnamefont
  {Kubo}} \emph {et~al.},\ }\href@noop {} {\emph {\bibinfo {title} {Statistical
  Physics II: Nonequilibrium Statistical Physics}}},\ \bibinfo {edition} {2nd}\
  ed.\ (\bibinfo  {publisher} {Springer-Verlag},\ \bibinfo {address}
  {Heidelberg},\ \bibinfo {year} {1995})\BibitemShut {NoStop}%
\bibitem [{\citenamefont {San~Miguel}\ and\ \citenamefont
  {Chaturvedi}(1980)}]{san1980limit}%
  \BibitemOpen
  \bibfield  {author} {\bibinfo {author} {\bibfnamefont {M.}~\bibnamefont
  {San~Miguel}}\ and\ \bibinfo {author} {\bibfnamefont {S.}~\bibnamefont
  {Chaturvedi}},\ }\href {\doibase 10.1007/BF01295086} {\bibfield  {journal}
  {\bibinfo  {journal} {Z. Physik B}\ }\textbf {\bibinfo {volume} {40}},\
  \bibinfo {pages} {167} (\bibinfo {year} {1980})}\BibitemShut {NoStop}%
\bibitem [{\citenamefont {Yuan}\ \emph {et~al.}(2010)\citenamefont {Yuan} \emph
  {et~al.}}]{Yuan2010}%
  \BibitemOpen
  \bibfield  {author} {\bibinfo {author} {\bibfnamefont {R.-S.}\ \bibnamefont
  {Yuan}} \emph {et~al.},\ }\href@noop {} {\enquote {\bibinfo {title}
  {Constructive proof of global {Lyapunov} function as potential function},}\ }
  (\bibinfo {year} {2010}),\ \bibinfo {note} {arXiv:1012.2721}\BibitemShut
  {NoStop}%
\end{thebibliography}%

\end{document}